\magnification=1200
\hsize=16 true cm
\tolerance 2200

\vsize=23.5 true cm
\nopagenumbers      
\topskip=1 truecm
\normalbaselineskip=12pt
\normalbaselines
\headline={\tenrm\hfil\folio\hfil}
\raggedbottom
\abovedisplayskip=3mm
\belowdisplayskip=3mm
\abovedisplayshortskip=1mm
\belowdisplayshortskip=2mm
\def\ref#1{$[#1]$}
\vglue 2.5 true cm
\centerline{\bf {Universality in Random Systems: the case of 
 the 3-d Random Field Ising model} }
\vglue 1.5 true cm
\centerline{{\bf  Nicolas Sourlas}} 
\medskip
\centerline{Laboratoire de Physique Th\'eorique de l' Ecole Normale 
Sup\'erieure\footnote {*}  
{ Unit\'e Propre du Centre National de la Recherche Scientifique, 
associ\'ee \`a l' Ecole Normale Sup\'erieure et \`a l'Universit\'e de 
Paris-Sud. } }
\centerline{ 24 rue Lhomond, 75231 Paris CEDEX 05, France. }
\vglue 2.5 true cm
\centerline{\bf ABSTRACT}
\vglue 1 true cm

We study numerically the zero temperature Random Field Ising Model on 
cubic lattices 
of various linear sizes $ 6 \le L \le 90 $ in three dimensions with 
the purpose of verifying the validity of universality for disordered 
systems. 
For each random field configuration we vary the ferromagnetic coupling 
strength $J$ and compute the ground state {\it exactly}.
 We examine the case of different random field probability 
distributions: gaussian distribution, zero width  bimodal distribution $h_{i} = \pm 1$, 
 wide  bimodal distribution $h_{i} = \pm 1 +\delta h$ (with a gaussian $\delta h$). 
We also study the case of the randomly-diluted antiferromagnet in a field, which is 
thought to be in the same universality class. 
 We find that in the infinite volume limit the magnetization 
is discontinuous in $J$ and we compute the relevant exponent, which,
 according to finite size scaling, equals $ 1/ \nu $ . 
We find different values of $ \nu $ for the different random field 
distributions, in disagreement with universality.

\vglue 1.5 true cm
\noindent LPTENS 98/33
\medskip
\noindent September 98

\vfill\eject

It is well known in the theory of phase transitions that critical exponents 
of very different physical systems can be identical. This property is
 called universality. Universality is one of the most important 
concepts in modern physics and can be explained in the framework  
of the perturbative renormalization group, where it is shown that 
critical exponents depend only on very few ``relevant" parameters of 
 the physical system (dimentionality of space, dimentionality of the 
order parameter, symmetry properties, ...) 
and do not depend on the details of the interactions. 

It is generally assumed that universality is also valid in the presence 
of quenched disorder. The theoretical justification requires, in this 
case, the use of the 
replica trick, in addition to the perturbative renormalization group(PRG), 
and is therefore weaker in the case of disordered systems. 
Furthermore it is known that PRG fails in the case of the random 
field Ising model (RFIM)\ref{1}. 
 The RFIM is (together with branched polymers\ref{2}) 
 one of the very few cases where PRG can be
 analyzed to all orders of pertubation theory\ref{3,4} and this analysis leeds to 
false conclusions (while for the branched polymers the same analysis 
is correct!). 

Universality is difficult to check because the difference of the 
critical exponents in different systems is in general small. 
 Different attempts to check universality in disordered Potts models 
or diluted ferromagnets 
in two dimensions gave conflicting results. 
The purpose of this paper is to check universality in the case of the RFIM 
in three dimensions using numerical simulations. In a previous paper\ref{5} to 
be referred in the following as {\bf I }, we studied the cases of the gaussian 
and bimodal distribution of random fields and 
we found evidence that universality is  violated in the RFIM. In this paper 
we present also the cases of the diluted random antiferromagnet in a 
field (DRAF) which is believed to be in the same universality class and of 
the wide bimodal distribution (see below). 
The evidence for violation of universality becomes much stronger.

 We would like to mention the very recent  preprint  
by Ballesteros et al.\ref{6} on three dimentional diluted Ising ferromagnets. 
Theese autors measure the critical exponents and universal ratios for 
different dilutions. Their results are dilution independent, in agreement with 
universality, provided that 
they take into account the first nonleading correction 
to scaling.  

The Hamiltonian of the RFIM is given by 

$$ H \  = \ -J \sum_{<i,j>} \ \sigma_{i} \sigma_{j} \ - \  \sum_{i} \ 
\ h_{i} \sigma_{i} \eqno(1) $$
where $ \sum_{<i,j>} $ runs over neighbouring sites of the lattice 
(we have only considered three dimensional cubic lattices with periodic 
boundary conditions) and 
$ h_{i} $ are independent random variables identically distributed with 
probability distribution $ P(h) $.  In the following we will consider the case 
of zero mean gaussian distribution $  P(h) =  {1 \over{ \sqrt{2 \pi } } } 
{ \exp  {-h^{2}/2 } }   $,
of the bimodal distribution $  P(h) = {1 \over 2 } (\delta (h-1) + \delta (h+1) ) $, 
and of the wide bimodal distribution $  P(h) = {1 \over { 2 w \sqrt{2 \pi } } } ( 
\exp { -(h-1)^{2}/2 w^{2} }   + \exp { -(h+1)^{2}/2  w^{2} }   ) $.
We also considered the case of the deleted random antiferromagnet in a 
field $H$ which is believed to be in the same universality class. When the 
ratio $ H/J $ is rational, additional singularities appear in the distribution 
of the ground states. For this reason we considered a random field in this 
case also, with probability distribution $  P(h) =  {1 \over { w \sqrt{2 \pi } } }
\exp { -(h-1)^{2}/2 w^{2} }  $. It is straitforward to show 
in the framework of PRG that 
the addition of a random component does not change the universality class. 
We considered the case of site dilution with probability $ p=.40 $ and w=.1

For a given random field sample, one can vary both 
the ferromagnetic coupling $J$ and the temperature $T$, i.e. the phase 
boundary is a line in the $J$, $T$ plane. It is thought, in accordance 
with PRG,  that the nature of 
the transition and the value of the exponents do not depend on the position 
 on the transition line, nor on the direction one crosses  it,
 and that this is true down to zero temperature. Furthermore it has 
been argued that the 
renormalization group flow drives the system to $ T=0 $, i.e that 
 $ T=0 $ is an attractive fixed point of the 
renormalization group.
So it is advantageous to work at $T=0$ where it has been shown that 
the RFIM is equivalent to the problem of maximum flow in a graph\ref{7},
 for which 
 very fast (polynomial) algorithms are known. Another advantage of this algorithm 
is that it provides the exact groud state and therefore there is no 
thermalization problem. Simulations using such 
algorithms have already 
been performed in the past\ref{8,9}. In the present paper we use the
 latest version of 
the algorithm developped by Goldberg and Tarjan\ref{10},
 which we optimized for the case of 
the cubic lattice. It has been shown that this algorithm converges to the 
ground state in a time $t < L^{6} \ln L $ where L is the linear size of the 
cubic lattice. We found experimentally that $ t \sim L^{4} $\ref{11}.

We proceeded as follows: We first produced a configuration of the external fields.
 We then chose $k$ values of the ferromagnetic coupling $J_1,J_2,...,J_k $
and we computed the corresponding ground states. 
We considered lattices of different linear sizes $L$, from 
 $L = 6 $ to $L=90$. We studied between 1000 samples for  $L=90$ 
and  40000 samples for $L=6$.
We studied the variation of the absolute value of the magnetization $ m(J) $ and 
of the energy as a function of $J$. 
We found that there is a region in $J$ where there are large discontinuities 
of $ |m| $ and that outside that region $ m(J) $ is a smooth function of $J$. 
The amplitude of the discontinuities is volume independent, while the 
width of the region in $J$ where they appear, shrinks as the volume increases. 
 For every $  \{ h_{i} \} $ sample, we chose the 
$n_{d}$ largest variations of the 
magnetization between two succesive values of $J $. Let's call 
$j_1 < j_2 < \cdots < j_{n_{d}} $ the values of $J$ at which they occur.
 The choice of $n_{d}$ is somehow arbitrary. We took $2 \le n_{d} \le 6 $. 
Our results are compatible with the hypothesis

$$ j_{i} \ = \ j_{\infty} \ + \ { c_{i} \over L^{p} } 
\ (1 \ + \ { f_{i} \over L^{q} } \ )  + \delta j_{i} \eqno(2) $$
$ \delta j_{i} $ are zero mean gaussian random variables with a width 
$ \sigma $ decreasing with the volume 

$$ \sigma (L) \sim  \sigma_{0} L^{- \delta}  ( 1 + \sigma_{1} 
 L^{- \rho } )  \eqno(3)  $$
 As the volume increases 
the mean is shifted towards the infinite volume value $ j_{\infty} $, 
which is the same for all the $ j_{i} $'s and at the same time the 
variance of the distributions shrinks to zero. This 
shift of the ``critical coupling" is analogous to the shift 
of the effective critical temperature due to finite volume 
corrections, well known to occur in ordinary second order phase 
transitions. 
We conclude that the appearance of several discontinuities in the 
magnetization is a finite volume artifact and that our results are fully
 compatible with the hypothesis of a single discontinuity in the 
thermodynamic limit. 
This is true for 
all distributions of the random field we studied. 
Figure 1 shows the average absolute value of the magnetization difference 
$ Dm_2 $ as a function of $ 1/L $ for 
the case of the DRAF and for the wide bimodal 
distribution and $ Dm_4 $ for the 
gaussian case, where 
$ Dm_k = \overline { | m(j_k)^{+} -m(j_1)^{-} | }  $,
 $ m(j_k)^{+} = \lim_{j \to j_k , \  j >  j_k } m(j) $ and 
 $ m(j_1)^{-} = \lim_{j \to j_1 , \  j <  j_1 } m(j) $, i.e.  $ Dm_k $
is the average magnetization difference computed  between  the $k$ largest 
 discontinuities.

These discontinuities have not been seen in the previous simulations for the 
following reason. Only the average magnetization has been measured and 
for finite volumes, the position of the dicontinuities 
fluctuates from sample to sample so that the average magnetization 
is continuous. 
This raises the question of how to average over the disorder. The 
only method available for analytic calculations is the replica method. 
With this method one can compute  the averages of observables 
for fixed values of the couplings. This in turn has inspired the 
numerical work. 
 But there is more freedom with numerical 
simulations, as we have illustrated above. 
We see in the present case that averaging the magnetization for  fixed values  
of the ferromagnetic coupling, hides the  true nature of the transition, 
i.e. the discontinuity of the magnetization.

In figure 2 we plot  
 $ Dj =  \overline  {j_2 - j_1 }  $, i.e. the average 
difference of the values of the couplings 
at which the two largest discontinuities of the magnetization occur, 
versus $ 1/L $. The continuous lines are the best fit to the data. 
The upper line is for the DRAF and the lower line for the wide bimodal 
distribution. The interested reader can find the data for the gaussian 
distribution in {\bf I }. We conclude that the ansatz of equation (4)
describes well the data down to a value of $L$ as small as $L=6$.
To be more quantitatif, the $ \chi^{2} $
 per degree of freedom is $ \chi^{2} =.9  $ for the gaussian case, 
$ \chi^{2} =1.1  $ for the DRAF and $ \chi^{2} =2.75  $ for the 
wide bimodal distribution.

Let's now discuss some possible origins of systematic errors. 
An obvious one is the choice of $ n_d $, the number of discontinuities 
we choose to analyse. The number of large discontinuities fluctuates from 
sample to sample and the choice  of $ n_d $ is not obvious. 
We have choosen the largest value of $ n_d $ which is compatible with 
the data, i.e. provides a $ \chi^{2} $
 per degree of freedom of the order of one. This is $ n_d = 5 $ for the 
gaussian distribution and $ n_d = 2 $ for the other distributions. 
We studied the dependence of the exponents on $ n_d $ 
and we found it to be small. 
 A detailed discussion for the gaussian distribution 
is found in {\bf I }. 
 Another important point is the choice of the values of  
the ferromagnetic couplings $J_1 < J_2, \cdots , < J_m $ at which we compute 
the ground states. The values $ j_i $ at which the strong discontinuities 
of the magnetization occur, strongly fluctuate from sample to sample. 
So if the total range of $ J_1 $ to $ J_m $ is not large enough, we may 
miss some of the discontinuities. On the other hand the  $ J_k $'s 
must be dense enough in order to be able to observe a discontinuity. 
The obvious way to satisfy both requirements is to compute a very large 
number of ground states per sample but this is impossible because of computer 
time limitations. For the gaussian distribution (see {\bf I } ) 
we chose equally spaced $x = 1/J $'s with $\delta x = .0125 $ 
and required that  for every size $ L $, $ J_1 < { \overline j_1 } - 4 \sigma_1 $ 
where $ { \overline j_1 } $ is the average location of the first sigularity 
for that size and $ \sigma_1 $ its variance. We required similarly 
that $ J_m > { \overline j_n } + 4 \sigma_n $ where $ j_n $ is the last 
singularity. For the wide bimodal distribution and the DRAF we 
proceeded differently. Let the ground state energy for 
the coupling $ J $ be $ - E (J) $ and the corresponding value of the 
``exchange energy" be $ E_e (J) = \sum_{<i,j>} \ \sigma_{i} \sigma_{j} $.
It can be shown that $ E (J) $ is a convex function of $ J $. 
If we know $ E (J) $ and its derivative $ E_e (J) $ for  
 $ J_1 < J_2 \cdots < J_k $ then we have, because of the convexity,
 both an upper and 
a lower bound for $ E (J) $ for $ J_1 < J < J_k $. We start by 
computing a few ground states (small $ k $) sparsely sampling a very 
large $ J $ region. Then we compute the values of $ J $ at which the 
difference of the 
upper and the lower bound is maximum and we compute the 
ground states for those values of the $ J $. We iterate the procedure 
until the difference between the upper and the lower bound is everywhere 
smaller than $ 5 \times  10^{-6} \times E(J) $. We further require that 
the resolution in $ J $ around the stronger discontinuities be better than 
$ 10^{-4} $. This is an efficient way to cover a large domain in $ J$,
while concentrating computational effort where $ E (J) $ 
 and $ m(J) $ vary the most. 

Despite the magnetization discontinuity, we do not think that this is a 
first order phase transition for two reasons:  
the non-classical values of the exponents and the continuity of the 
energy derivative. It was shown in {\bf I}  
that the discontinuities of the energy derivative, which appear at 
the same value of the coupling as the magnetization discontinuities, 
vanish in the infinite volume limit. In the case of the bimodal distribution, 
additional energy discontinuities appear at rational values of 
the couplings. These discontinuities do not disappear in the infinite volume limit.
 We found that they disappear if we add a small width 
to the field distribution. A similar phenomenon is observed for the DRAF. 
This is the reason we 
introduced the wide bimodal distribution and we added a 
gaussian width to the field in the diluted antiferromagnet case.

 Figure 3 shows the values of the exponents $ p$ and $ q $ that 
are compatible with our data (within 90\% confidence) 
for the different random field distributions.  
We see that the values of 
$ p $ and $q$ are correlated and the uncertainties on $p$ and $q$ are 
quite large. 
 Previous simulations were unable to determine the first correction to 
scaling, because of the limited number of sizes and of 
 statistical errors, so the data were fitted to a single power law. 
We found that the two different fitting assumptions,  a single power law behaviour  or the inclusion of a subleading correction, lead to quite different values 
of the exponent $ 1 / \nu $ and drive it considerably out of the statistical 
error bars. For a detailed discussion and comparison with previous work see 
{\bf I}. 

We also determined the exponents $ \delta $ and $ \rho $ which 
measure sample to sample fluctuations (see equation   (3)).
We find that these exponents are identical to $ p $ and $ q $ 
within our error bars.  
A hand waiving argument to explain this is to say that  
the appearence of several singularities, the shift 
of the critical value of the coupling and the sample to sample fluctuations 
are finite size effects which are controlled by the ``dimensions of the 
same operators". We feel that this argument has to be substanciated. 
We find this result remarkable and consider it as a confirmation of our 
analysis. 

According to finite size scaling, $ p = 1/ \nu $  and $ q $ 
should  be 
universal, i.e. the same for all probability distributions of the random 
field. We see that our data are compatible with the hypothesis 
that the random diluted 
antiferromagnet in a field and the gaussian random field are in the 
same universality class. This is a highly nontrivial result of 
perturbative renormalization group, because 
the two systems are completely different. The bimodal and wide 
bimodal distributions are also mutually compatible for $ q > 3 $. 
This is not visible in figure 3. But if one takes into account all 
the probability distributions of the random fields our data 
are clearly not compatible with universality,  despite the large 
uncertainties on $ p $ and $ q $. 
This is also confirmed  by the following  
analysis. We can fit the data for the three random field distributions, 
i.e. the gaussian, the wide bimodal and DRAF,  
 with the assumption of universality, i.e. the assumption that they share 
the same values of the exponents.  
 We then find  $ p = .55 $, 
$ q = 1.25 $ with a $ \chi^{2} $
 per degree of freedom  $ \chi^{2} = 11.2  $ This is to be compared 
with  $ \chi^{2} = 1.6  $ per degree of freedom for the hypothesis 
of different exponents. We conclude that our data present strong evidence 
for the violation of universality in the RFIM at zero temperature, 
contrary to what is commonly believed. 

One may argue that a second nonleading correction to finite site scaling may 
restore universality. We do not know of any theoretical argument excluding 
such corrections. But if this additional term is necessary for the  
determination  of  critical exponents in disordered systems, 
 then it would be impossible to measure  them numerically (and probably 
also experimentally) in the near future. 

After completion of this work we became aware of a preprint by 
Hartmann and Nowak\ref{12} where they study numerically the 
ground state properties of random field Ising models in three dimensions.

I would like to thank J.-C. Angl\`es d'Auriac for his pleasant and fruitfull 
collaboration in the first part of this work.

This paper is dedicated to Heinz Horner on the occasion 
of his 60th birthday.


\vglue 1 true cm


\centerline{\bf References }
\vglue .6 true cm

\item {1)} For recent reviews see:
T. Natterman and J. Villain, {\it Phase Trans. } {\bf 11} 817 (1988)
 \hfill\break
D. P. Belanger and A. P. Young, {\it J. Magn. Magn. Mater } {\bf 100 }
 272 (1991)
\item {2)} G. Parisi and N. Sourlas, { \it Phys. Rev. Lett. }
 {\bf 46 } 871 (1981)
\item {3)} A. P. Young, {\it J.Phys. A} {\bf 10} L257 (1977)
\item {4)} G. Parisi and N. Sourlas,{ \it Phys. Rev. Lett. }
 {\bf 43 } 744 (1979)
\item {5)} J.-C. Angl\`es d' Auriac and N. Sourlas, {\it Europhys. Lett.} 
{\bf 39 } 473 (1987)
\item { 6)} H. G. Ballesteros, L. A. Fernandez, V. Martin-Mayor, A. Munoz Sudupe,
 G. Parisi and J. J. Ruiz-Lorenzo, cond-mat/9802273 
\item {7)} J.-C. Angl\`es d' Auriac, M. Preismann and R. Rammal, 
{\it J. de Physique-Lett. } {\bf 46} L173 (1985)
\item {8)} J.-C. Angl\`es d' Auriac, Thesis, Grenoble (1986)
\item {9)} A. T. Ogielski {\it Phys. Rev. Lett. }, {\bf 57 } 
1251 (1986) 
\item {10)} A. V. Goldberg and R. E. Tarjan {\it Journal of the Association for  
Computing Machinery }  {\bf 35} 921 (1988)
\item {11)} J.-C. Angl\`es d' Auriac, M. Preismann and A. Sebo, {\it J. of 
Math. and Computer Modelling}, to appear.
\item {12) } A. K. Hartmann and U. Nowak, cond-mat/9807131


\vfill\eject

\vglue .7 true cm
\centerline{\bf Figure captions }
\item { Figure 1.} Average magnetization discontinuities as a function of  
inverse lattice size $ 1/L $. Diamonds show the data for the 
 gaussian distribution, crosses for the wide bimodal distribution and 
circles for the randomly diluted antiferromagnet in a field.

\item { Figure 2.} Average 
difference of the values of the couplings 
for the two largest discontinuities of the magnetization, 
as a function of  
inverse lattice size $ 1/L $. Diamonds show the data for the  
wide bimodal distribution and 
circles for the randomly diluted antiferromagnet in a field.

\item { Figure 3.} 90{\%} confidence levels for the exponents $p$ and $q$ 
 for the different random field distributions.
G denotes the gaussian distribution, AF the diluted antiferromagnet in a 
field, BM the mimodal distribution and WBM the wide bimodal distribution.


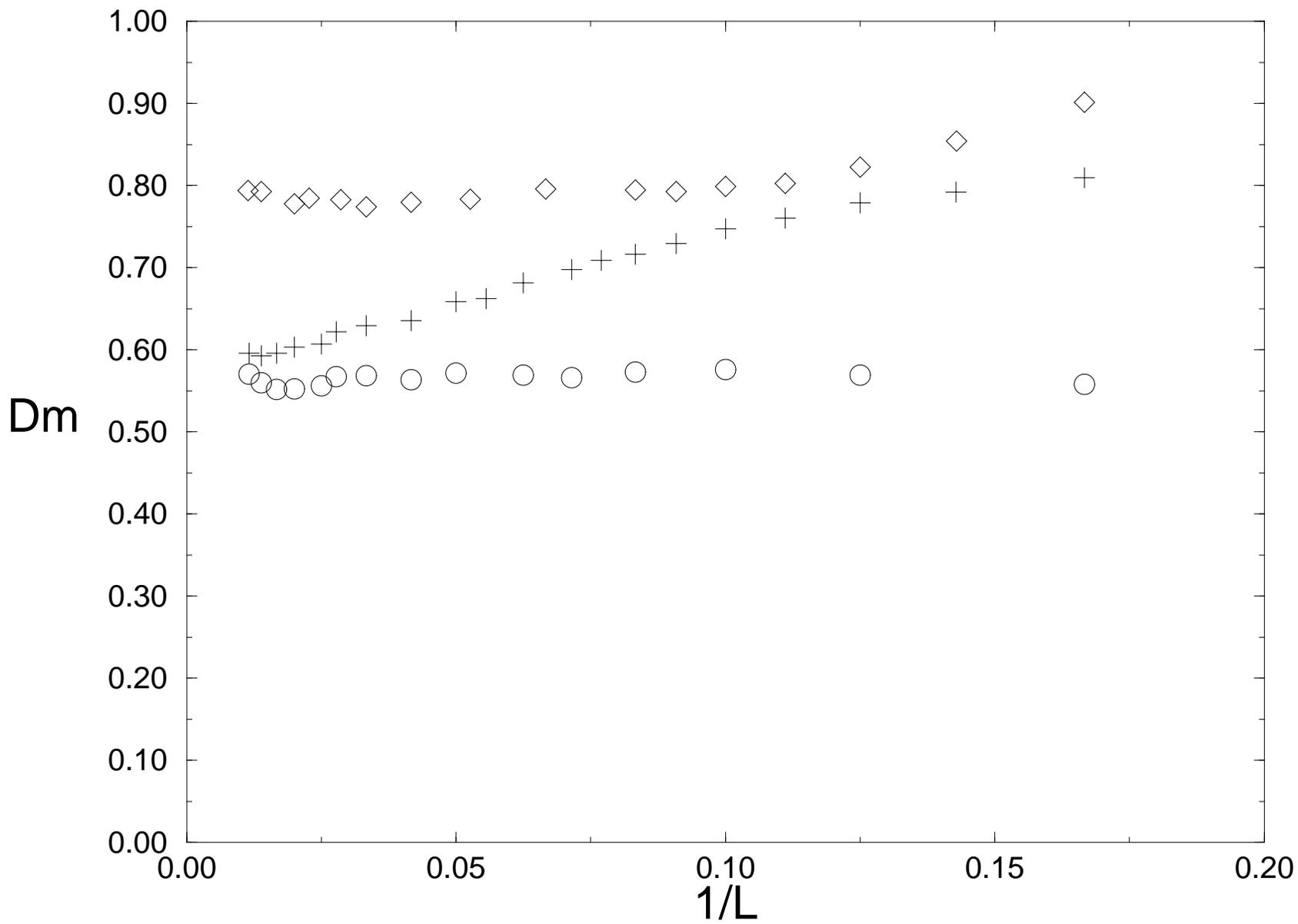
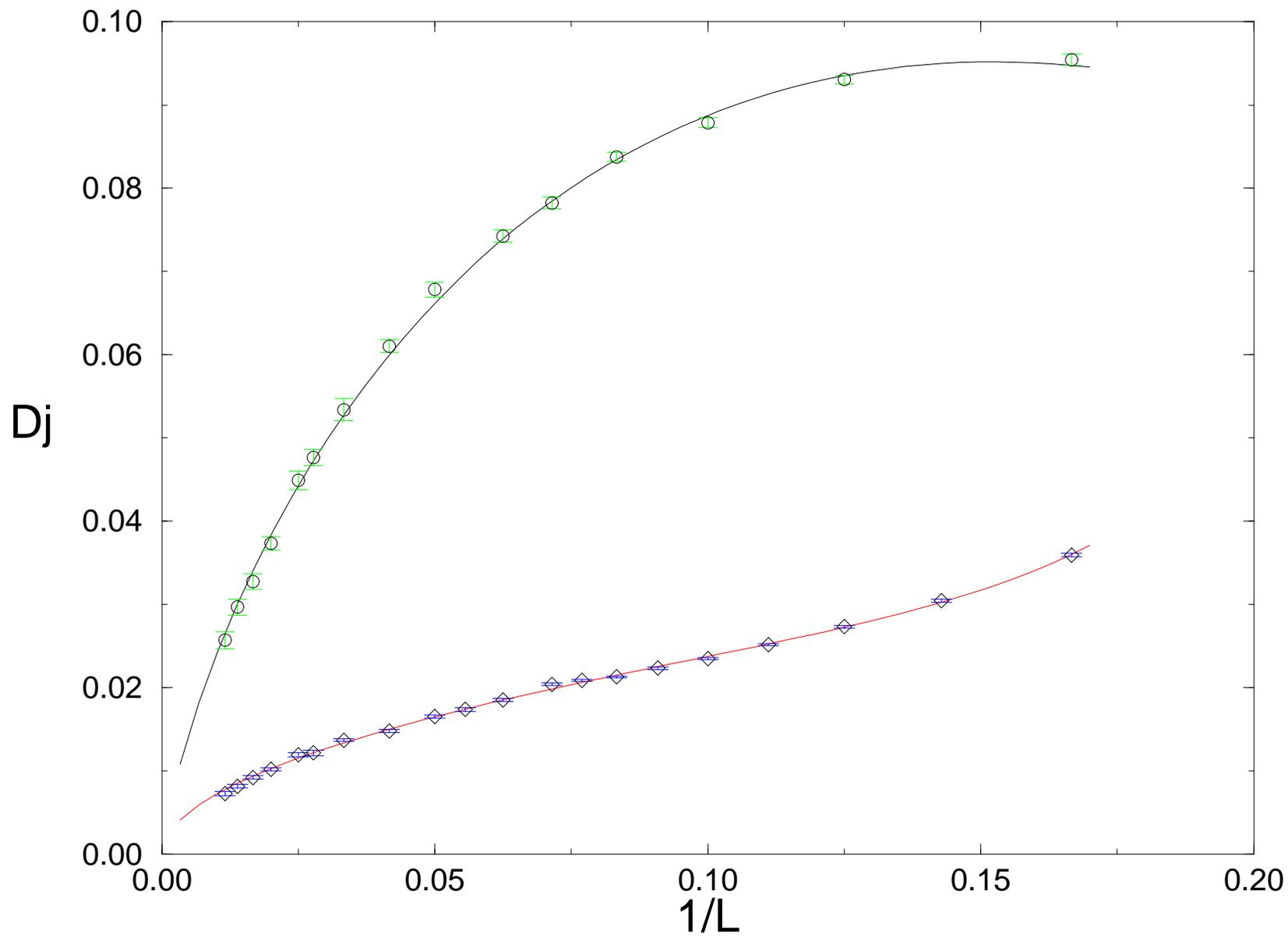
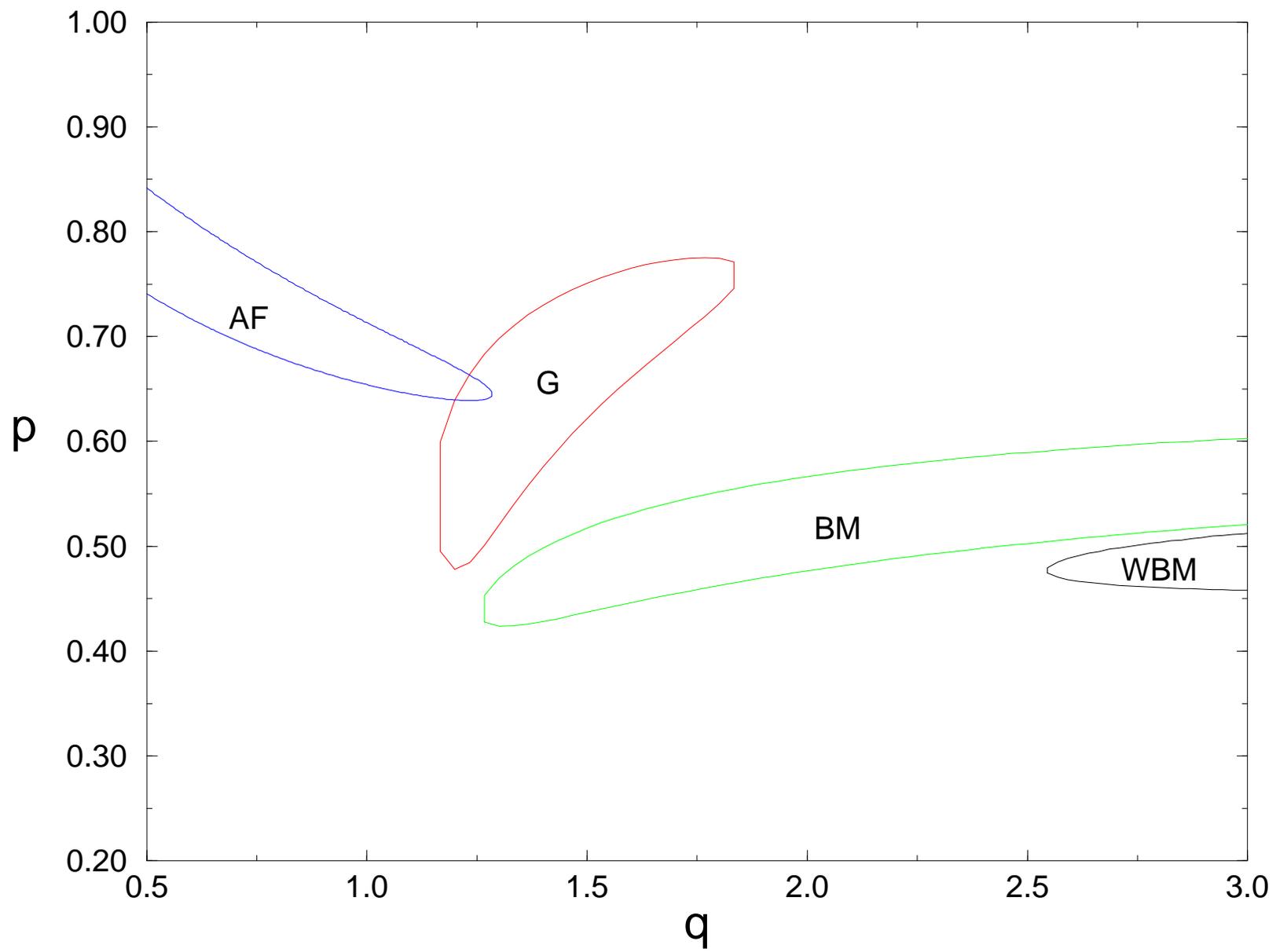
\end